\documentclass[sn-mathphys-num]{sn-jnl}

\usepackage{graphicx}%
\usepackage{multirow}%
\usepackage{amsmath,amssymb,amsfonts}%
\usepackage{amsthm}%
\usepackage{mathrsfs}%
\usepackage[title]{appendix}%
\usepackage{xcolor}%
\usepackage{textcomp}%
\usepackage{manyfoot}%
\usepackage{booktabs}%
\usepackage{algorithm}%
\usepackage{algorithmicx}%
\usepackage{algpseudocode}%
\usepackage{listings}%
\usepackage{array}%

\theoremstyle{thmstyleone}%
\theoremstyle{thmstyletwo}%

\theoremstyle{thmstylethree}%

\begin{document}

\title[The Impact of Quantum-Safe Cryptography (QSC) on Website Response]{The Impact of Quantum-Safe Cryptography (QSC) on Website Response}

\author*[1,2]{\fnm{Ananya} \sur{Tadepalli}}\email{atadepalli31@gatech.edu}





\abstract{Modern web traffic relies on 2048-bit RSA encryption to secure our data in transit. Rapid advances in Quantum Computing pose a grave challenge by allowing hackers to break this encryption in hours. In August of 2024, the National Institute of Standards and Technology published Quantum-Safe Cryptography (QSC) standards, including CRYSTALS-Kyber for general encryption and CRYSTALS-Dilithium, FALCON, and SPHINCS+ for digital signatures. Despite this proactive approach, the slow adoption of encryption protocols remains a concern, leaving a significant portion of data vulnerable to interception. In this context, this study aims to evaluate the impact of NIST's Quantum-Resistant Cryptographic Algorithms on website response times, particularly focusing on SSL handshake time and total download time under varying network conditions. By assessing the performance of these algorithms, this research seeks to provide empirical evidence and a reusable framework for validating the efficacy of QSC in real-world scenarios. It was found that the QSC algorithms outperformed the classical algorithm under normal and congested network conditions. There was also found to be an improvement in the total download time for larger file sizes, and a better performance by QSC under higher latency and packet loss conditions. Therefore, this study recommends that websites switch to QSC when the standards are ratified. These insights are crucial for accelerating the adoption of QSC and ensuring the security of data in the face of quantum computing threats. }





\maketitle

\section{Introduction}\label{sec1}
 In less than five years, scientists believe that quantum computers will be able to crack current public encryption systems, threatening the confidentiality of bank accounts, financial markets, and vital infrastructure \cite{herman2021qday}. Present data is secured using the RSA 2048-bit encryption scheme—a public-key system where the sender and receiver need to exchange a secret sequence of bits called a key, which is kept private—that works through the factorization of a large prime number \cite{panhwar2021quantum} and would take a cyber-criminal 300 trillion years to decrypt \cite{herman2021qday}. However, quantum computers, a rapidly emerging technology that harnesses the laws of quantum mechanics to solve problems too complex for classical computers \cite{ibm2023quantum} will be able to decrypt this encryption scheme in less than 8 hours \cite{herman2021qday}. Therefore, the development and ratification of an encryption algorithm that can protect data from exposure by Quantum computers is a prevalent initiative that has been undertaken by the National Institute of Standards and Technology (NIST) and the cybersecurity industry. Four of these algorithms, commonly referred to as Quantum-Safe Cryptography (QSC) or Post-Quantum Cryptography algorithms, were officially selected by NIST: CRYSTALS-Kyber for general encryption and, CRYSTALS-Dilithium, FALCON, and SPHINCS+ for digital signatures \cite{tlsgigamon2023}. The crux of the issue lies in the slow adoption of encryption protocols despite the widespread recognition of their necessity in safeguarding data during transit. Even 25 years after the introduction of the TLS standard, only 85 percent of data in transit is encrypted, leaving a substantial portion vulnerable to interception \cite{tlsgigamon2023}. Compounding this concern is the escalating sophistication of hackers in data acquisition, posing a significant threat to digital security. It is critical that websites adopt these QSC algorithms following their official ratification to secure data before the commercial arrival of Quantum computers. To accelerate and champion the adoption of these QSC algorithms, this study poses and answers the question, “What is the impact of the National Institute of Standards and Technology’s (NIST) Quantum-Resistant Cryptographic Algorithms for Digital Signatures, when integrated with the algorithm for General Encryption, on Website Response Time in the context of SSL Handshake Time and Total Download Time under varying network conditions?”
Developing on this question, this study hopes to evaluate the algorithms for digital signatures and how they affect their website performance. The security of encryption schemes that will protect data in the era of quantum computing depends on both the developers of the algorithms and the users who adopt them. As NIST and other Quantum researchers continue to evaluate the security offered by these algorithms, research analyzing how they will impact secure Transport Layer Security (TLS) communication—the security protocol that provides privacy and data integrity for Internet communications \cite{nist2022quantum} —is important to the user-side responsibility of securing data against cyber-criminals.


\section{Literature Review}\label{sec2}

\subsection{Context}\label{subsec21}

For the purpose of this study, it is important to understand the historical and mathematical development of classical algorithms and the quantum algorithms that can break them. 
Public-key cryptographic systems, such as RSA 2048-bit encryption, are based on computational complexity \cite{sikeridis2020post}. A “difficult” algorithm exhibits computational complexity because the time a computer takes to perform a specific task will grow exponentially as the number of bits in the input increases. RSA algorithms are based on the factorization of large prime integers meaning that as the digit of the integer grows, the time it takes to factorize it increases exponentially \cite{sikeridis2020post,gisin2002quantum}. This is called the Integer Factorization Problem (IFP), and it is the basis of data security in classical computing \cite{kumar2022post}. 
In 1994, Peter Shor developed Shor’s algorithm which can solve the IFP in polynomial time with the use of a quantum computer \cite{kumar2022post}. With Shor’s algorithm, a quantum computer with sufficient number of qubits, a basic unit of quantum information, and fault tolerance will be able to crack RSA 2048-bit encryptions \cite{gisin2002quantum}.

\subsubsection{QSC Algorithms}\label{subsec22}
\subsubsection{CRYSTALS-KYBER}\label{subsubsec221}

The first of the four algorithms that has been ratified by NIST is the CRYSTALS-Kyber (p384-kyber768) algorithm. According to NIST, this algorithm is a key encapsulation mechanism (KEM) \cite{borges2020comparison}, which means under certain conditions, it can be used by two parties to establish a shared secret key over a public channel \cite{das2022challenges}. KYBER generates public and private keys, encapsulates the shared secret key using the recipient's public key, and allows for secure key exchange. This algorithm is considered secure based on the theoretical foundation of lattice-based cryptography \cite{borges2020comparison}. It is designed to be resistant to various attacks, and there is evidence to support its security claims.

\subsubsection{CRYSTALS-Dilithium}\label{subsubsec222}

Digital signatures are used to verify identities during a digital transaction or to sign a document remotely The first of these three algorithms is CRYSTALS-Dilthium (dilithium3) which is based on the Fiat-Shamir paradigm \cite{borges2020comparison}. The Fiat-Shamir paradigm is a method for secure identification and signature without relying on traditional public or shared keys. It utilizes the IFP to maintain security and is designed to be provably secure against known or chosen message attacks \cite{alagici2022nist}. Dilithium is built upon the "Fiat-Shamir with aborts" approach, which employs a three-message lattice-based identification scheme. This scheme allows a prover to convince a verifier of their possession of the secret key (s1, s2) without revealing it \cite{borges2020comparison}.

\subsubsection{FALCON}\label{subsubsec223}

The second algorithm for digital signatures is FALCON (falcon512), Fast Fourier Lattice-based Compact Signatures over NTRU, which is a lattice-based signature scheme utilizing the “hash-and-sign” paradigm. It generates a public-private key pair, where the public key is derived from a set of polynomials. Signing with FALCON involves a complex trapdoor preimage sampling algorithm, and its security is based on the hardness of the Short Integer Solution (SIS) Problem over NTRU lattices. FALCON offers small signature sizes and fast verification, making it suitable for constrained protocol scenarios \cite{borges2020comparison}.

\subsubsection{SPHINCS+}\label{subsubsec224}
The third and final algorithm for digital signatures is SPHINCS+ (sphincssha2128fsimple). SPHINCS+ is a unique and innovative stateless hash-based signature scheme that brings together several cryptographic techniques, including one-time signatures, few-times signatures, Merkle trees, and hypertrees, to create a versatile digital signature system suitable for a wide range of applications \cite{borges2020comparison}. SPHINCS+ offers varying security levels, with parameters chosen to provide 128, 192, or 256 bits of security. Unlike some other signature algorithms, SPHINCS+ is stateless, which makes it more robust \cite{borges2020comparison, moody2023module}.

\subsection{Societal Impact}\label{subsec23}

Researchers have found that the use of single photons and polarization in quantum cryptography makes it impossible for eavesdroppers to copy or modify encrypted messages transmitted through optical fiber channels \cite{fiat1987how}, providing “unconditional security”. This unconditional security is achieved through a method called Quantum Key Distribution (QKD) QKD is a cryptographic technique that leverages the principles of quantum mechanics to create secure communication channels. The first QKD protocol, BB84, was developed in 1984 and used the polarization states of single photons to distribute cryptographic keys securely. This is referred to as discrete variable (DV) QKD \cite{aumasson2019sphincs}. 
When comparing this to classical computing to establish the treat that quantum computers pose, it is found that in traditional public-key cryptography, trapdoor functions are used to conceal the meaning of messages between two users from a passive eavesdropper, despite the lack of any initial shared secret information between the two users. In quantum public key distribution, the quantum channel is not used directly to send meaningful messages, but is rather used to transmit a supply of random bits between two users who share no secret information initially, in such a way that the users, by subsequent consultation over an ordinary non-quantum channel subject to passive eavesdropping, can tell with high probability whether the original quantum transmission has been disturbed in transit, as it would be by an eavesdropper \cite{muruganantham2020quantum}.
\begin{wrapfigure}{r}{0.4\textwidth}
  \begin{center}
    \includegraphics[width=0.4\textwidth]{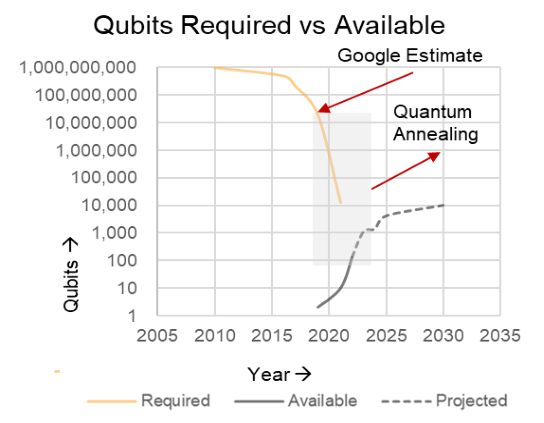}
  \end{center}
  \caption{Graphical presentation of projected stable qubit availability vs number of stable Qubits required} \label{Figure 1}
\end{wrapfigure}
\subsection{Q-Day}\label{subsec24}
Q-Day is defined as the day when quantum computers will be commercially available. Unlike Y2K, Q-Day is not a specific date and estimates are changing as development of stable Q-bits is progressing. Figure 1 below demonstrates the data based on current information \cite{herman2021qday}.
Quantum annealing models, which combine a gate-based approach with an optimization approach \cite{scientific2024prime}, are further reducing stable Qubit requirements by combining quantum and classical algorithms. With uncertainty surrounding Q-Day and ever reducing cos of data storage, there is a new concern around sophisticated hackers storing today’s data in the hope of cracking them in the future. Buoyed by this concern companies such as Cloudflare are already switching to quantum safe cryptography.

\subsection{Transport Layer Security}\label{subsec25}

Unconditional security becomes possible by utilizing QKD as proven by various research. This implies that security in this context remains independent of the eavesdropper's capabilities, and it is found that technological advancements will not pose a threat to security. Therefore, researchers propose to integrate QKD with the TLS protocol \cite{muruganantham2020quantum}.
However, researchers propose significant modifications to the TLS Handshake Protocol in order to enhance security. The primary objective is to establish security parameters through QKD and eliminate reliance on PKI (Public Key Infrastructure). This is achieved by sharing a secret (S) between the client and server and replacing traditional key exchange methods (e.g., RSA or Diffie-Hellman) with the BB84 protocol for QKD. To address BB84's vulnerability to "man in the middle" attacks, the protocol checks for eavesdropping by calculating the TLS finished messages on both the client and server sides using the shared secret S and the key K derived from the BB84 Protocol \cite{muruganantham2020quantum}.
In accordance with these modifications, studies suggest that over 60 percent of Internet connections are implemented over the TLS-based secure HTTPS protocol \cite{djordjevic2022physical}. TLS adoption is expected to keep increasing as users and client vendors strive for ubiquitous encryption and privacy \cite{elboukhari2010improving, chan2018monitoring}.
The performance of QSC on TLS secure communication has been tested. One study provides a comprehensive performance assessment of NIST signature algorithm candidates, considering realistic network conditions and their impact on TLS 1.3 connection establishment latency. It also examines the effect of these algorithms on a server's achievable TLS session throughput, exploring the trade-off between longer QSC signatures and computationally intensive QSC operations for both idle and heavily loaded servers. The results of this study indicate that adopting at least two QSC signature algorithms is a viable option for time-sensitive applications over TLS, with minimal additional overhead compared to current signature algorithms \cite{chan2018monitoring}. This study provides a basis for further research to measure the impact of QSC on TLS secure communication with varying network conditions.
There is a stigma towards encryption in transit in web community due to additional time it takes to establish secure connection and transmit data. After 50 years of TLS, 15 percent of web traffic still flows over http. This statistic will play an important role in challenges we will face for switching to QSC. This project would provide both empirical evidence and a repeatable framework to load test different websites before and after switch.

\subsection{Gap}\label{subsec26}
The necessity for websites to adopt QSC to people’s privacy and security presents a compelling need to accelerate its adoption. Previous research in this area provides insight into the development of these QSC algorithms and their expected capability to rebuild the cryptographic vault, protecting it against both quantum and classical attacks \cite{borges2020comparison}. 
Then, with the incorporation of QSC over secure TLS communication channels, researchers find that modifications to this protocol must be made to ensure a smooth transition to QSC algorithms \cite{muruganantham2020quantum}. The network conditions that will impact the QSC’s ability to establish TLS connection has been studied in a narrow capacity with the focus on sever load and latency time. Specifically, previous research has developed a framework, that will be utilized in this study’s methods, for evaluating the performance of post-quantum cryptographic primitives TLS protocol under realistic network conditions but focuses on the testing of this framework on strictly latency and packet loss with little variability \cite{sikeridis2020post}. 
The current research does not answer two key questions in the minds of website owners. First, what will be the impact of this algorithm on my customers and second how will the impact be in different network and load conditions. After 30 years of https, we notice 21\% of top 100,000 websites still server http content \cite{tlsgigamon2023, ssl2021https}.
This study aims to build a reusable framework for testing individual website performance under different network conditions to broaden the record evidence of how switching to QSC will impact the performance. Availability of empirical evidence and a reusable framework for validation is expected to accelerate the adoption of QSC.

\section{Methods}\label{sec3}

To address the posed research question, this study will design an experiment to assess the impact of the National Institute of Standards and Technology's (NIST) First Four Quantum-Resistant Cryptographic Algorithms (QSC) on website response time. Experimental, quantitative research was chosen to answer the posed research question because the impact of QSC can best be observed by observing the effects of varying network conditions (independent variables) on the website response (dependent variables) as introduced by similar studies in this field \cite{kotzias2018coming}. The researcher hypothesizes that the adoption of QSC will not lead to more than 20\% increase in total website response time.
The impact of QSC on website response will be measured by using a combination approved Quantum-safe key exchange algorithms (p384 kyber768) and Quantum-safe digital signature algorithms (dilithium3, falcon512, sphincssha2128fsimple) against a classical key exchange algorithm (X25519) and a classical digital signature algorithm (ed25519).
The experimental setup will utilize a Docker engine running on an AWS cloud server, providing a controlled and scalable environment for testing. The test environment will feature a test server configured with QSC and a test client similarly configured. Additionally, a separate test server will be configured with standard Transport Layer Security (TLS) encryption. The environment will be designed to support various network scenarios to allow this study to independently control variables such as latency and packet loss rate, and then examine the performance impact of various post-quantum primitives on TLS connection establishment. Open Quantum Safe, an open-source community \cite{oqs2023home}, offered Docker containers specifically designed for this type of research \cite{sikeridis2020post}. Docker container images from Docker Hub, will be modified for this study including the Quantum Safe Server, Quantum Safe Client, Traffic Controller (for introducing different network latencies and packet losses), and a Classic Server for performance comparison. This design model was initialized and tested by “Benchmarking Post-Quantum Cryptography in TLS”; however, that study focused solely on how varying packet loss affects post-quantum algorithms that fragment across many packets \cite{kotzias2018coming}.

\begin{wrapfigure}{r}{0.4\textwidth}
  \begin{center}
    \includegraphics[width=0.4\textwidth]{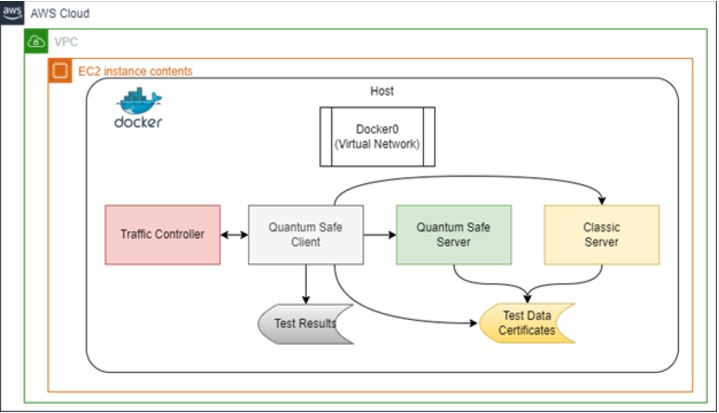}
  \end{center}
  \caption{Method Architecture} \label{Figure 2}
\end{wrapfigure}

This study will focus on analyzing two critical dependent variables: SSL Handshake Time and Total Download Time. SSL Handshake Time will represent the duration required to establish a secure SSL connection, while Total Download Time will indicate the overall time taken to retrieve web content (measured in seconds). These dependent variables will be examined in the context of three independent variables: File Size, Latency, and Packet Loss. File Size will reflect variations in the size of files being transferred (measured in MB), Latency will quantify the delay in data transmission (seconds), and Packet Loss will measure the rate at which data packets are lost during transit (expressed as a percentage of packets that were sent but not received). The examination of these independent variables will aim to discern how alterations in network conditions affect the crucial aspects of website response, shedding light on the potential impact of QSC on the performance of websites under diverse network scenarios. 
As outlined in Figure 2, this study will utilize a 2MB file as the base file size for the purpose of performance comparison, drawing from research on average web transfer size \cite{ori2023experimental}. Three network scenarios, as outlined through internet research \cite{paquin2020benchmarking, pingplotter2022latency, kang2022importance}, will be identified for testing purposes, including the Ideal Network with zero latency and zero packet loss, the Normal Network with 100ms latency and 1.5\% packet loss, and the Congested Network with 200ms latency and 2.5\% packet loss. This study hypothesizes that there will be a less that 10\% increase in SSL Handshake Time and Total Download Time under these varying network conditions.
This study will explore various additional scenarios to assess sensitivity to alterations in file size, latency, and packet loss. These scenarios will involve adjusting file sizes in factors of 2, ranging from  244 bytes to 16.3GB. Latency will be varied from 0 to 400ms in increments of 50 milliseconds, and packet loss will be adjusted from 0\% to 5\% in increments of 0.5.

\begin{table}[h]
\caption{Caption text}\label{tab1}%
\begin{tabular}{>{\raggedright\arraybackslash}p{0.4\linewidth}>{\raggedright\arraybackslash}p{0.1\linewidth}>{\raggedright\arraybackslash}p{0.07\linewidth}>{\raggedright\arraybackslash}p{0.06\linewidth}>{\raggedright\arraybackslash}p{0.05\linewidth}}
\toprule
Scenario&  File Size&  Latency&  Packet Loss& Runs\\
\midrule
Impact under three operating scenarios (ideal, normal, and congested)&  2 MB&  0ms
100ms
 200ms&  0\%
1.5\%
2.5\%& 100
100
100\\
         Impact as file sizes changes &  244 bytes to 16 GB &  200 ms&  2.5\%& 297\\
         Impact as latency changes &  2 MB&  0 to 400ms in 50ms increments&  2.5\%& 88\\
         Impact as loss changes &  2 MB&  200 ms&  0 \% to 5\% in increments of .5\%& 121\\
\botrule
\end{tabular}
\footnotetext{Source: This is an example of table footnote. This is an example of table footnote.}
\caption{Scenarios for data collection}
\label{Table 1}
\end{table}

In this study, the following methodology will be employed to assess the performance of quantum-safe (qs) servers in comparison to classic servers on an Amazon Web Services (AWS) environment. First, research credits will be obtained from AWS for cloud usage. Subsequently, a test server will be constructed using the AWS console. The Docker engine and Docker Compose will be installed to facilitate containerized deployment. Local Docker images for qs-server, qs-client, and classic-server will be constructed based on the provided base images, with key modifications such as reducing TLS caching settings to 2 seconds to allow for multiple tests without caching and enabling NETADMIN for traffic control. Various sizes of test files will be generated using Linux commands. Docker Compose files will be created for each scenario, configuring file and SSL key sharing, network settings, and creating a shared folder for result reporting. A shell script will be devised to run multiple tests and record pertinent time measurements in a CSV file (see Appendix A). Containers for each test scenario will be spun up, and tests will be executed on the qs-client. Finally, the test server will be dismantled. This comprehensive methodology ensures a systematic approach to evaluating the performance of quantum-safe servers against their classic counterparts in a controlled AWS environment.

\section{Results}\label{sec4}
\subsection{Network Conditions}\label{subsec41}
\subsubsection{Ideal}\label{subsubsec411}

Under ideal network conditions of 0 ms of latency and 0\% packet loss, the average SSL Handshake Time was found for each of the four scenarios: the classical algorithm, kyber/dilithium, kyber/falcon, and kyber/sphincsha. Figure 3 (left) illustrates the distribution of SSL Handshake Time values under the simulated ideal network conditions using box plots for each cryptographic algorithm tested. Within each box plot, the central line represents the median SSL Handshake Time, while the box itself spans the interquartile range (IQR), encapsulating the middle 50\% of the data. The upper and lower whiskers extend to the highest and lowest data points within 1.5 times the IQR from the upper and lower quartiles, respectively, serving as indicators of the data's spread. It was found that under ideal network conditions, the mean SSL Handshake Time for the classical algorithm was approximately 0.01 sec in comparison to approximately 0.03 sec for kyber/dilithium, kyber/falcon, and kyber/sphincsha.
Figure 3 (right) shows the frequency distribution of the classical algorithm in comparison to the three variation of the quantum algorithms, kyber/dilithium, kyber/falcon, and kyper/sphincsha. The mean total download time for the classical algorithm was 0.0315 sec, while the mean for the kyber/dilithium algorithm was 0.0507 sec, the mean for the kyber/falcon algorithm was 0.0478 sec, and the mean for the kyber/sphincsha algorithm was 0.0468 sec. 
This demonstrates that under ideal network conditions, there was a >10\% in SSL Handshake Time and Total Download Time for every QSC algorithm when compared to the classical algorithm.

\begin{figure}
\includegraphics[width=\textwidth]{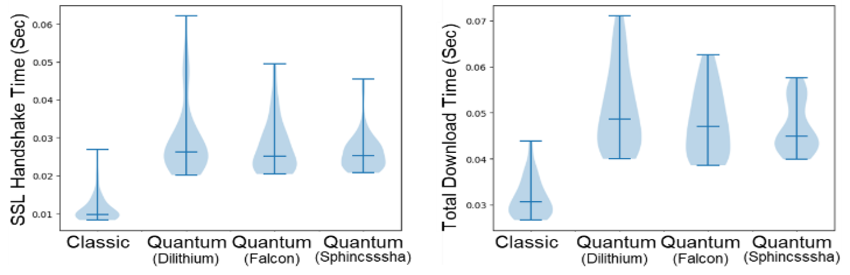}
\caption{Box Plot of SSL Handshake Time and Total Download Time values for the classical algorithm and three variations of QSC algorithms—kyber/dilithium, kyber/falcon, and kyber/sphincsha—under ideal network conditions.} \label{Figure 3}
\end{figure}

\subsubsection{Normal}\label{subsubsec412}

Under normal network conditions characterized by 100 ms of latency and 1.5\% packet loss, the average SSL Handshake Time was determined for each of the four scenarios: the classical algorithm, kyber/dilithium, kyber/falcon, and kyber/sphincsha. Figure 4 (left) illustrates the distribution of SSL Handshake Time values under the simulated normal network conditions using box plots for each cryptographic algorithm tested. It was found that under normal network conditions, the mean SSL Handshake Time for the classical algorithm was approximately 0.32 sec in comparison to approximately 0.23 sec for kyber/dilithium, 0.31 sec for kyber/falcon, and 0.31 sec for kyber/sphincsha. Therefore, the mean SSL Handshake Time decreased by approximately 28.13\% when comparing classical cryptography to Kyber/dilithium, with decreases of 3.13\% observed for both Kyber/falcon and Kyber/sphincs+ compared to classical cryptography.
Similarly, the mean total download time for the classical algorithm and the three variations of QSC algorithms was assessed under normal network conditions. Figure 4 (right) presents the frequency distribution of total download time values for the classical algorithm and the three variations of QSC algorithms—kyber/dilithium, kyber/falcon, and kyber/sphincsha. It was found that under normal network conditions, the mean Total Download Time for the classical algorithm was approximately 5.2234 sec in comparison to approximately 5.1424 sec for kyber/dilithium, 5.4973 sec for kyber/falcon, and 4.2771 sec for kyber/sphincsha. This accounts for a 1.55\% decrease, 5.24\% increase, and 19.07\% decrease between the classical algorithm and Kyber/Dilithium, Kyber/Falcon, and Kyber/Sphincs+, respectively.
This demonstrates that under ideal network conditions, there was a <10\% in SSL Handshake Time and Total Download Time for every QSC algorithm when compared to the classical algorithm.

\begin{figure}
\includegraphics[width=\textwidth]{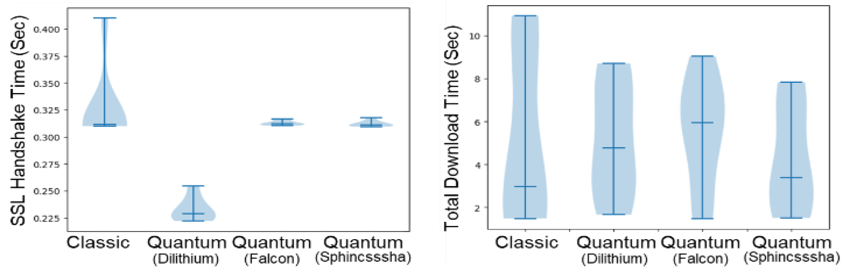}
\caption{Box Plot of SSL Handshake Time and Total Download Time values for the classical algorithm and three variations of QSC algorithms—kyber/dilithium, kyber/falcon, and kyber/sphincsha—under ideal network conditions.} \label{Figure 4}
\end{figure}

\subsubsection{Congested}\label{subsubsec413}

In contrast to ideal network conditions, the network environment was deliberately congested to simulate real-world scenarios with elevated latency (200 ms) and increased packet loss rates (2.5\%). Figure 5 (left) depicts the distribution of SSL Handshake Time values for the classical algorithm and the three variations of QSC algorithms—kyber/dilithium, kyber/falcon, and kyber/sphincsha—under congested network conditions. 
It was observed that under congested network conditions, the mean SSL Handshake Time for the classical algorithm exhibited a substantial increase compared to ideal conditions, reaching approximately 0.77 sec. Similarly, the QSC algorithms experienced elevated mean SSL Handshake Times, with kyber/dilithium, kyber/falcon, and kyber/sphincsha averaging approximately 0.42 sec, 0.61 sec, and 0.64 sec, respectively. This accounts for a a 45.45\% decrease in SSL Handshake Time between classical and Kyber/Dilithium, a 20.78\% decrease between classical and Kyber/Falcon, and a 16.88\% decrease between classical and Kyber/Sphincs+
The same result was observed in Figure 5 (right) for Total Download Time which averaged approximately 23.5496 sec for the classical algorithm and 20.9333 sec, 18.3020 sec, and 16.8645 sec for kyber/dilithium, kyber/falcon, and kyber/sphincsha, respectively. Specifically, this provides that there was a 11.11\%, 22.28\%, 28.39\% decrease in Total Download Time between the classical algorithm and Kyber/Dilithium, Kyber/Falcon, and Kyber/Sphincs+, respectively.
This demonstrates that under congested network conditions, there was a <10\% in SSL Handshake Time and Total Download Time for every QSC algorithm when compared to the classical algorithm. This indicates that network congestion adversely impacts the efficiency of SSL handshake procedures across all cryptographic algorithms tested and that quantum algorithms perform better under more network congestion.

\begin{figure}
\includegraphics[width=\textwidth]{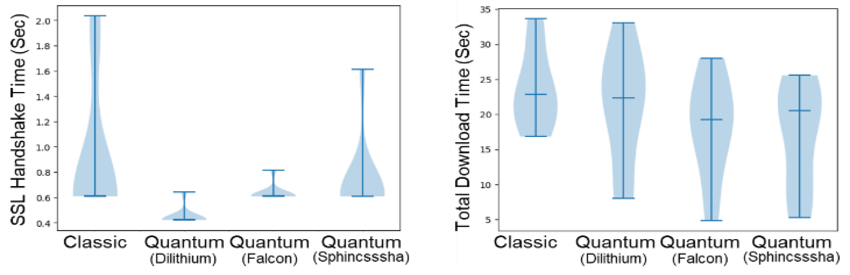}
\caption{Box Plot of SSL Handshake Time and Total Download Time values for the classical algorithm and three variations of QSC algorithms—kyber/dilithium, kyber/falcon, and kyber/sphincsha—under congested network conditions.} \label{Figure 5}
\end{figure}

\subsubsection{Sensitivity Analysis}\label{subsec42}
\subsubsection{File Size}\label{subsubsec421}

Figure 6 shows the distribution of SSL Handshake Time values for each cryptographic algorithm tested, including the classical algorithm and the three variations of the quantum algorithms: kyber/dilithium, kyber/falcon, and kyber/sphincsha, when latency and packet loss were held constant with varying file sizes. It was found that the quantum algorithms exhibited higher SSL handshake times than the classical algorithms when latency and packet loss were kept constant.

\begin{figure}
\includegraphics[width=\textwidth]{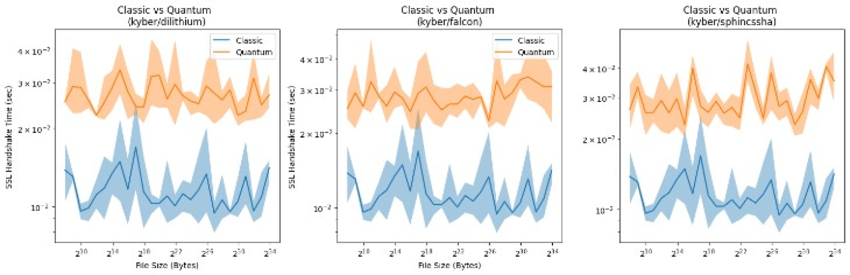}
\caption{Distribution of SSL Handshake Time values for classical and QSC algorithms under varying file sizes.} \label{Figure 6}
\end{figure}

Figure 7 shows the total download time as file size is varied from 244 bytes to 20 GB. The graph illustrates the relationship between file size and total download time for each cryptographic algorithm tested, including the classical algorithm and the three variations of the quantum algorithms: kyber/dilithium, kyber/falcon, and kyber/sphincsha. It was found that the quantum algorithms, on average, had higher total download times than the classical algorithm when latency and packet loss were held constant.

\begin{figure}
\includegraphics[width=\textwidth]{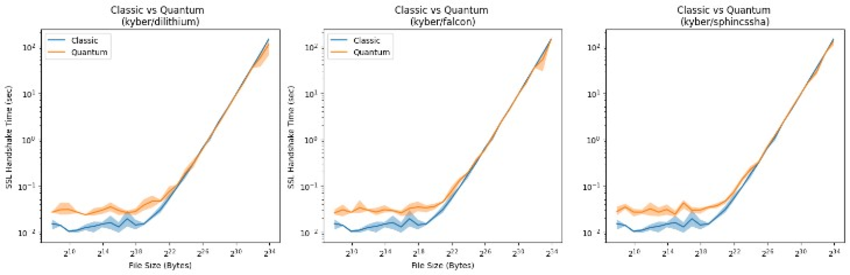}
\caption{Distribution of Total Download Time values for classical and QSC algorithms under varying file sizes.} \label{Figure 7}
\end{figure}

\textbf{Byte Transfer Rate}: In addition to assessing total download time, the byte transfer rate serves as a critical performance metric for evaluating the efficiency of cryptographic algorithms in data transmission. In figure 8, three separate graphs depict the byte transfer rate of the classical algorithm compared to each variation of the quantum algorithms—kyber/dilithium, kyber/falcon, and kyber/sphincsha—as file size is systematically increased. It was found that at the file size was increased from 244 bytes to 20 GB, the difference in byte transfer rate between the classical and the three quantum algorithms became more negligible.
The results in Figure 8 indicate that the quantum algorithms had higher SSL Handshake Times and Total Download Times that the classical algorithm.

\begin{figure}
\includegraphics[width=\textwidth]{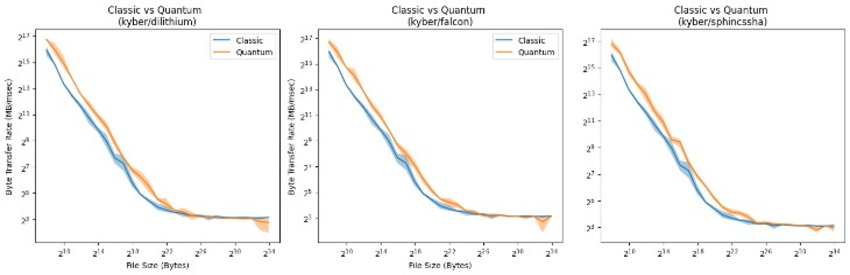}
\caption{Distribution of Byte Transfer Rate values for classical and Quantum-Safe Cryptography (QSC) algorithms under varying file sizes.} \label{Figure 8}
\end{figure}

\subsubsection{Latency}\label{subsubsec422}

Figure 9 illustrates the impact of latency on SSL Handshake Time for the classical algorithm and the three variations of Quantum-Safe Cryptography (QSC) algorithms—kyber/dilithium, kyber/falcon, and kyber/sphincsha. As latency increases from 10 ms to 500 ms, the SSL Handshake Time for all cryptographic algorithms also increases. However, the quantum algorithms more frequently exhibit lower SSL Handshake Times compared to the classical algorithm across varying levels of latency.

\begin{figure}
\includegraphics[width=\textwidth]{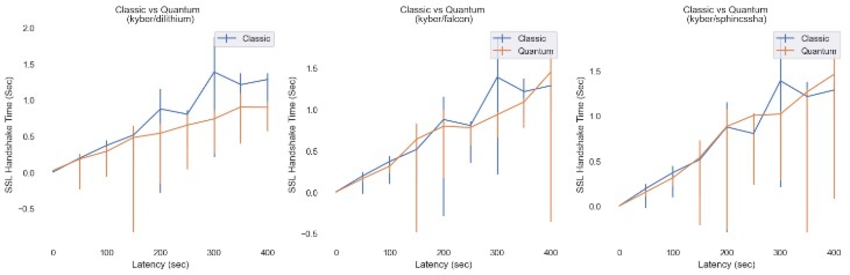}
\caption{Distribution of SSL Handshake Time values for classical and Quantum-Safe Cryptography (QSC) algorithms under varying latency values} \label{Figure 9}
\end{figure}

Figure 10 displays the corresponding Total Download Time values as latency is varied from 10 ms to 500 ms. The graph demonstrates that as latency increases, the Total Download Time for all algorithms also increases. Similarly, the quantum algorithms tend to have higher Total Download Times than the classical algorithm across different latency levels, albeit with some variability.

\begin{figure}
\includegraphics[width=\textwidth]{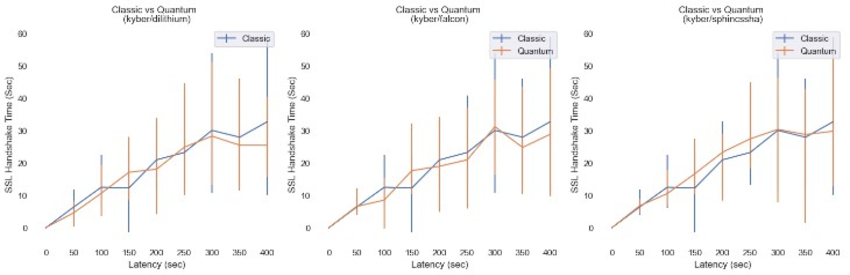}
\caption{Distribution of Total Download Time values for classical and Quantum-Safe Cryptography (QSC) algorithms under varying latency values} \label{Figure 10}
\end{figure}

\subsubsection{Packet Loss}\label{subsubsec423}

Examining the effect of packet loss on SSL Handshake Time, Figure 10 presents the SSL Handshake Time values for the classical algorithm and the three quantum algorithm variations under varying packet loss rates from 0\% to 5\%. It is observed that as packet loss increases, the SSL Handshake Time for all algorithms increases. Notably, the quantum algorithms consistently exhibit lower SSL Handshake Times compared to the classical algorithm across different levels of packet loss.

\begin{figure}
\includegraphics[width=\textwidth]{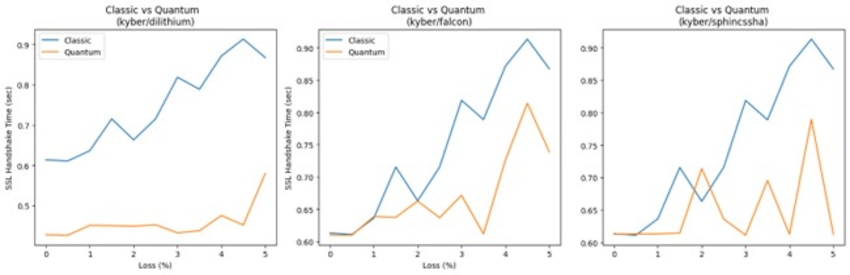}
\caption{Distribution of SSL Handshake Time values for classical and Quantum-Safe Cryptography (QSC) algorithms under varying packet loss percentages} \label{Figure 11}
\end{figure}

Figure 12 depicts the Total Download Time values corresponding to different packet loss rates ranging from 0\% to 5\%. As packet loss rates increase, the Total Download Time for all cryptographic algorithms also increases. It was observed that the quantum algorithms tend to have higher Total Download Times than the classical algorithm across different latency levels, albeit with some variability.

\begin{figure}
\includegraphics[width=\textwidth]{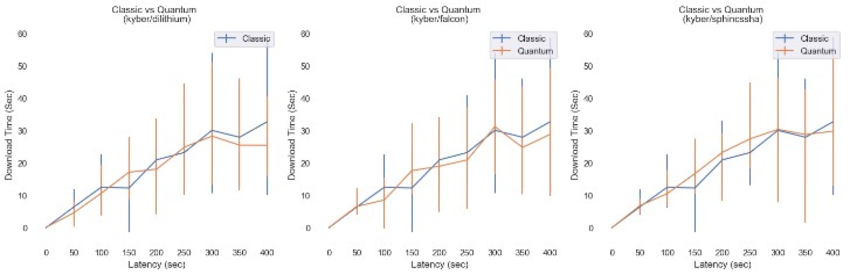}
\caption{Distribution of Total Download Time values for classical and Quantum-Safe Cryptography (QSC) algorithms under varying packet loss percentages} \label{Figure 12}
\end{figure}

\section{Discussion}\label{sec5}

The experiment's findings offer insights into the behavior of classical and QSC algorithms across diverse network conditions to answer the posed research question of “What is the impact of the National Institute of Standards and Technology’s (NIST) Quantum-Resistant Cryptographic Algorithms for Digital Signatures, when integrated with the algorithm for General Encryption, on Website Response Time in the context of SSL Handshake Time and Total Download Time under varying network conditions?” 
From the three scenarios tested-ideal, normal, and congested-it was found that under ideal network conditions, the hypothesis that “There will be a less that 10\% increase in SSL Handshake Time and Total Download Time between the classical and QSC algorithms under varying network conditions” was not supported. However, under normal and congested network conditions, this hypothesis was supported, actually revealing a reduction in total website response time when transitioning from classical to QSC algorithms. However, one notable exception emerged in a specific scenario: the total download time exhibited an increase of 5.24\% under normal network conditions when transitioning from classical to Kyber/Falcon encryption, but this remains below the 10\% threshold that was hypothesized.
In extrapolating the implications of this experiment to real-world scenarios concerning the adoption of Quantum-Safe Cryptography (QSC) algorithms, several critical insights emerge. It is essential to acknowledge the inherent impracticality of achieving ideal network conditions in real-world networks, where complete absence of packet loss and latency approaching zero milliseconds is unattainable. Given this reality, the demonstrated superior performance of Quantum algorithms compared to classical counterparts under both normal and congested network conditions carries significant weight. This study offers compelling evidence advocating for the widespread adoption of QSC algorithms by websites as a proactive measure to safeguard sensitive data from potential exposure to quantum computers in the future.
The sensitivity analysis conducted in this experiment revealed crucial insights into the behavior of classical and QSC algorithms across various network conditions. Initially, under ideal network circumstances characterized by minimal latency and packet loss, the SSL Handshake Time and Total Download Time remained relatively consistent across all cryptographic algorithms tested. However, deviations from this ideal scenario led to significant variations in performance. This information can have real-world applications as websites can utilize the sensitivity analysis results to observe the impact of switching to QSC based on the network conditions clients experience on their networks.
The findings suggest that QSC algorithms offer competitive performance compared to classical algorithms across a range of scenarios, including ideal, normal, and congested network conditions. Despite the computational complexity inherent in QSC algorithms, they demonstrate resilience to adverse network conditions and maintain secure connections with minimal performance degradation.

\section{Conclusion}\label{sec6}

These findings have significant implications for the imminent arrival of quantum computer and the threat they pose to current classical RSA encryptions. By demonstrating the feasibility and efficacy of QSC algorithms in diverse network environments, this study contributes to the ongoing discussion surrounding the adoption of quantum-resistant cryptographic standards. The observed performance improvements underscore the importance of websites switching to QSC algorithms as viable alternatives to classical algorithms to protect consumer data. This study presents a unique approach to this pressing issue by providing empirical evidence of the impact of switching to QSC in the context of varying network conditions (file size, packet loss, and latency) in order to examine website response. 
The findings of this study provide evidence of the resilience of Quantum-Safe Cryptography (QSC) algorithms in maintaining secure connections across diverse network scenarios. While the ideal network conditions exhibited relatively stable performance across all cryptographic algorithms tested, deviations from this scenario, such as increased latency and packet loss, revealed significant differences in performance. Under normal and congested network conditions, where latency and packet loss were heightened, the quantum algorithms showcased improved performance compared to classical algorithms. Despite the inherent computational complexity of QSC algorithms, they demonstrated lower mean SSL Handshake Times and Total Download Times, indicating their suitability for securing data transmission in adverse network environments.

\section{Limitations}\label{sec7}

However, it is essential to acknowledge the limitations of this study, including the controlled nature of the experimental setup and the simplified network conditions simulated. Real-world network environments may exhibit additional complexities and variations not captured in this study, which could affect the generalizability of the findings. While the outlined experimental design and methodology provided a comprehensive framework for assessing the impact of Quantum-Safe Cryptography (QSC) on website response time, it is important to acknowledge certain limitations inherent in this research approach. First, the reliability and accuracy of the findings were dependent on the functioning of the software components involved, particularly the Quantum Safe Server, Quantum Safe Client, Classic Server, and Traffic Controller sourced from Docker Hub. Any software-related issues, such as bugs, compatibility issues, or unexpected behaviors, could have introduced confounding variables and compromised the validity of the results. Additionally, variations in network conditions beyond those explicitly tested may have existed in real-world scenarios, potentially limiting the generalizability of the findings. Despite these limitations, efforts were made to address and document any encountered issues throughout the experimentation process to enhance the transparency and reliability of the study's outcomes. 
Future research should strive to replicate these experiments in more realistic settings by considering a broader range of network conditions and operational environments. This could involve conducting experiments in diverse network infrastructures, including different types of networks such as wireless, cellular, or satellite networks, as well as incorporating varying levels of network congestion and fluctuating traffic patterns. Moreover, exploring additional factors beyond latency and packet loss that may influence the performance of Quantum-Safe Cryptography (QSC) algorithms is essential. These factors could include hardware specifications, such as processor speed and memory capacity, as well as the impact of different types of web content and applications on cryptographic algorithm performance

\section{Acknowledgements}\label{sec9}

I would like to thank Dr. Erin-Elizabeth Durham for her invaluable guidance, support, and feedback throughout the course of this research. Her expertise and encouragement have been crucial to the development and success of this project.
I am also grateful to Dr. Brendan Saltaformaggio whose insights, suggestions, and camaraderie have been instrumental in shaping this work.

\bibliography{bibliography}  
\paragraph{}
\begin{wrapfigure}{l}{35mm}
    \includegraphics[width=1in]{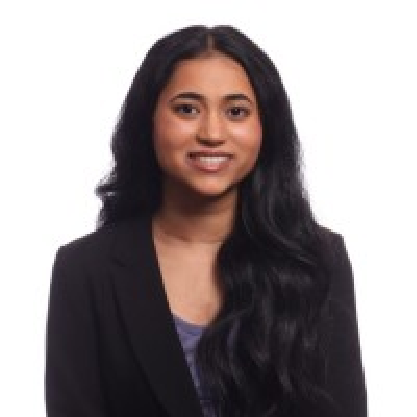}
     \label{ananya}
\end{wrapfigure}

\textbf{Ananya Tadepalli} 

\paragraph{}
A senior at Innovation Academy, Georgia with interests in research into Cybersecurity for quantum computing. She is currently interning at Georgia Institute of Technology Cyber Forensics Lab to develop her research skills.

\newpage

\begin{table}
    \centering
    \begin{tabular}{>{\raggedright\arraybackslash}p{0.05\linewidth}>{\raggedright\arraybackslash}p{0.85\linewidth}>{\raggedright\arraybackslash}p{0.05\linewidth}}
         Figure&  Caption&Page\\
         1&  Graphical presentation of projected stable qubit availability vs number of stable Qubits required&3\\
         2&  Method Architecture&5\\
         3&  Box Plot of SSL Handshake Time and Total Download Time values for the classical algorithm and three variations of QSC algorithms—kyber/dilithium, kyber/falcon,and kyber/sphincsha—under ideal network conditions&7\\
         4&  Box Plot of SSL Handshake Time and Total Download Time values for the classical algorithm and three variations of QSC algorithms—kyber/dilithium, kyber/falcon, and kyber/sphincsha—under ideal network conditions&7\\
 5& Box Plot of SSL Handshake Time and Total Download Time values for the classical algorithm and three variations of QSC algorithms—kyber/dilithium, kyber/falcon, and kyber/sphincsha—under congested network conditions&8\\
 6& Distribution of SSL Handshake Time values for classicaland QSC algorithms under varying file sizes&9\\
 7& Distribution of Total Download Time values for classical and QSC algorithms under varying file sizes&9\\
 8& Distribution of Byte Transfer Rate values for classical and Quantum-Safe Cryptography (QSC) algorithms under varying file sizes&10\\
 9& Distribution of SSL Handshake Time values for classical and Quantum-Safe Cryptography (QSC) algorithms under varying latency values&10\\
 10& Distribution of Total Download Time values for classical and Quantum-Safe Cryptography (QSC) algorithms under varying latency values&11\\
 11& Distribution of SSL Handshake Time values for classical and Quantum-Safe Cryptography (QSC) algorithms under varying packet loss percentages&11\\
 12& Distribution of Total Download Time values for classical and Quantum-Safe Cryptography (QSC) algorithms under varying packet loss percentages&12\\
    \end{tabular}
    \label{Table 2}
\end{table}

\end{document}